# Peer Evaluation of Video Lab Reports in an Introductory Physics MOOC

Shih-Yin Lin[#], Scott S. Douglas*, John M. Aiken*, Chien-Lin Liu, Edwin F. Greco*, Brian D. Thoms†, Marcos D. Caballero[¶], and Michael F. Schatz*

[#]*Department of Physics, National Changhua University of Education, Changhua, Taiwan 500*
*\*School of Physics, Georgia Institute of Technology, 830 State Street, Atlanta, GA 30332*
*† Department of Physics and Astronomy, Georgia State University, Atlanta, GA 30303*
*[¶] Department of Physics and Astronomy, Michigan State University, East Lansing, MI 48824*

**Abstract:** Assessing student performance becomes challenging when course enrollment becomes very large (~$10^4$ students). As part of a Massive Open Online Course (MOOC) in introductory physics offered by Georgia Tech in 2013, students submitted video reports on mechanics labs. Peer evaluation of these reports provided the primary method for evaluating student laboratory work. This paper describes the methods developed and used to guide students in evaluating each other's video lab reports. We also discuss how students' peer evaluation behavior changed with different interventions provided in the course.

## INTRODUCTION

As advancements in educational technologies have made distance learning more scalable, an increasing number of colleges and universities have started offering Massively Open Online Courses (MOOCs) [1]. However, assessing student performance in open online courses is challenging because course enrollment can be quite large (e.g., ~$10^4$ students). On assignments for which the computer-assisted grading capability is still limited, peer evaluation has become a frequently used method for assessing student work. Developing an effective peer evaluation structure to help students provide quality feedback to their peers therefore has become a paramount concern for the instructional design of these courses.

Since Summer 2013, Georgia Tech has been offering an introductory physics MOOC (open for students all over the world) in which students submit video reports on mechanics labs [2]. Peer evaluation is the primary method for evaluating student lab reports in the course. As research on how introductory physics students evaluate their peers' work [3] is currently limited, this study intends to explore students' evaluation behavior to help inform the design of an effective peer evaluation system. As a first step, we investigated what students attended to when evaluating each other's lab reports. In particular, we focused on students' evaluation of three lab reports assigned to all students in the MOOC, and investigated how these MOOC students took up different evaluative norms when different interventions were provided in subsequent semesters of the course.

## A MOOC with Laboratory Activities

We developed the Georgia Tech physics MOOC as an introductory calculus-based mechanics course [2]. Students in this MOOC experienced activities involving lectures (through online videos), textbook readings, homework assignments, exams, and laboratories. Students were able to interact with the instructors, TAs, and other students through an online discussion forum. In the Summer 2013 and Fall 2013 offerings, the course included four at-home laboratory activities that were designed to engage students in important scientific practices such as computational modeling and scientific communication. In each lab, students were instructed to observe an object undergoing a certain type of motion (e.g., constant velocity motion or falling motion under the influence of gravity and drag) and obtain position versus time data of the moving object (for example, by first capturing a video of the motion of the object, and analyzing the motion using the free video-tracking software Tracker [4]). Students then created computational simulations using VPython [5] to model and/or to explore the observed motion. Next, each student created a video lab report presenting their experiment, their model(s), and how the results of their experiment and model(s) related to each other. Finally, students uploaded their video lab reports to

| Item # | Item description | Grading options (and the corresponding point value) |
|---|---|---|
| 1 | **The video presentation is clear and easy to follow.** (Explanation: With this rubric question, feel free to criticize presentation, word usage, the quality of the video footage, and other features related to production value. After this question, please push production quality out of your mind and focus only on the topic the question is trying to probe.) | o **Strongly Agree** (12) o **Agree** (10) o **Neutral** (6) o **Disagree** (2) o **Strongly Disagree** (0) |
| 2 | **Does the video introduce the problem and state the main result?** (Explanation: Good introductions include a brief statement of the problem that's being addressed and the main punchline of the work that was done.) | o **Yes** (12) o **Hard to Tell** (6) o **No** (0) |
| 3 | **Does the video identify the model(s) relevant to this physical system?** (Explanation: Identifying the model(s) includes discussion of the main physics ideas (e.g., the momentum principle, curving motion), and how those ideas are applied in the problem under study.) | o **Yes** (12) o **Hard to Tell** (6) o **No** (0) |
| 4 | **The computational model(s) successfully predicts the motion of the object observed.*** (Explanation: When the computational model output and the observational data are plotted on the same graph, they match well with each other.) | o **Yes** (2) o **Hard to Tell** (1) o **No** (0) |
| 5 | **The presenter successfully discusses how his/her computational model(s) predicts or fails to predict the motion of the object.** (Explanation: The data used to initialize the model(s) are clearly identified. The parameters in the model that are adjusted to fit the data are clearly discussed. The places where the model fits the data well and where the model fits poorly are indicated.) | **Strongly Agree** (12) to **Strongly Disagree** (0) (Same as item 1) |
| 6 | **The video presentation correctly explains the physics.** (Explanation: With this rubric question, express your overall impression of how well the fundamental physics principles are connected to the motion under study) | **Strongly Agree** (12) to **Strongly Disagree** (0) (Same as item 1) |

*: In lab 3, item 4 is slightly changed to "The computational model(s) successfully predicts the mass of the black hole" to fit the purpose of the lab. All the other rubric items remain the same throughout all 4 labs.

**Table 1**. Peer evaluation rubric provided to the students.

YouTube, and submitted the links to their videos. These student-generated videos were later randomly distributed to other students for peer evaluation.

## PEER EVALUATION PROCESS

The rubric provided to students for use in peer evaluation is shown in Table 1. As enhancing students' communication skills and their understanding of how physics principles connect to the real world were central goals of our MOOC, this rubric focused on two major aspects of each video lab report: the quality of students' communication (i.e, Items 1 and 2 in Table 1) and the quality of students' discussion of the results and their explanation of the physics (i.e., Items 3-6 in Table 1). For each item, students were asked to rate the video report on a three-point scale (yes, hard to tell, no) or a five-point scale (strongly agree to strongly disagree). These scales were selected after we pilot tested the rubric to make sure it was easily interpreted for students. The point value of each grading option is listed in Table 1. In addition to these numerical ratings, students could also provide written suggestions in the text boxes following the rubric questions for how the presenter might improve on the evaluated item.

For each lab in the summer 2013 MOOC offering, every student was required to evaluate 5 video reports submitted by their peers (called the "peer videos") in the week immediately following video submission. Before the first peer evaluation assignment in Lab 1, students were instructed to watch one lecture video in which the instructors commented on two sample student video lab reports. Sample instructor evaluations using the rubric presented in Table 1 were also shown in the lecture video. Due to the limitations imposed by the course platform used in Summer 2013, no other intervention was provided for the students.

During the Fall 2013 offering, an additional intervention was provided; in each lab, students were required to grade four "practice videos" before they were allowed to evaluate their peers' work (four videos required). The practice videos were sample student lab reports selected from the previous offering of the course that spanned a wide range of quality in student work. After students submitted their evaluation of each practice video, they were shown the instructor ratings and comments on that same video for use as a reference. In order to provide incentives for students to read through the instructor evaluation and reflect on how their evaluation compared to the instructor evaluation, students were told that one-third of their laboratory score would depend on how closely their ratings aligned with instructor ratings on the third and fourth practice videos.

## DATA COLLECTION & METHODS

In order to gauge how students evaluate others' lab reports, in each lab, one of the peer videos assigned to each student was common among all students (hereafter called the "hidden calibration video"). The instructors evaluated the hidden calibration video so that a comparison between instructor evaluation and student evaluation could provide insight into how well students were taking up the evaluation norms when different interventions were provided. To avoid the potential problem of students grading the hidden calibration video differently than other peer videos, the hidden calibration appeared as a normal peer video when it was assigned to the students.

As the same hidden calibration videos were used between the first and second offering of the course in Labs 2, 3, and 4, we focus here on analyzing what students attended to in their evaluation of these three videos and how the intervention provided in the second run of the course affected students' evaluation behavior. Since many rubric items involve the evaluation of physics content in a video report, we wanted to ensure that students who participated in the lab activities in the summer and fall had similar physics backgrounds. The performance of the students on the Force and Motion Concept Evaluation (FMCE) [6] administered at the beginning of the course was therefore examined. There was no statistically significant difference on FMCE pre-test performance.

## RESULTS & DISCUSSION

Figure 1 shows the distribution of student ratings on each rubric item for the hidden calibration videos in Summer 2013. For each cell in Fig. 1, the white triangle indicates the instructors' rating on the same item in the same video. Comparison of students' ratings and instructors' ratings in Figure 1 suggests that in summer 2013, students had more lenient evaluation norms than instructors. For example, while instructors rated the Lab 4 video negatively on item 3 (*relevant physics principles*), 60% of the students rated it positively. Similar instances of a majority of students rating the video higher than instructors were consistently observed on other items in Lab 4. We note that if students had lenient evaluative norms, the agreement between student rating and instructor rating was likely to be high on well-produced videos that instructors also rated positively. As shown in Fig. 1, the student-instructor agreement was always higher than 60% on items 1 to 4 in Labs 2 and 3, where instructors' ratings were "Strongly Agree" or "Yes". When instructors' ratings were less positive (e.g., on item 5 *discussion of results* in Lab 2 and item 6 *overall physics discussion* in both Labs 2 and 3), the agreement became lower. Examining instructor comments on lab videos 2 and 3 suggests that both presenters seemed to display a fair understanding of the physics involved, except that the discussion was sometimes vague, involving small mistakes, or could have gone into more depth (e.g., saying "*The difference between the experiment and the model can be due to experimental error of force assumptions that were made*" without articulating the possible experimental errors). The fact that most students rated item 5 in Lab 2, item 6 in Lab 2, and item 6 in Lab 3 higher than instructors did suggest that these students were less likely to notice the nuanced flaws in other students' physics discussion, and/or that they might have a lower expectation of the physics quality required when evaluating others' lab reports.

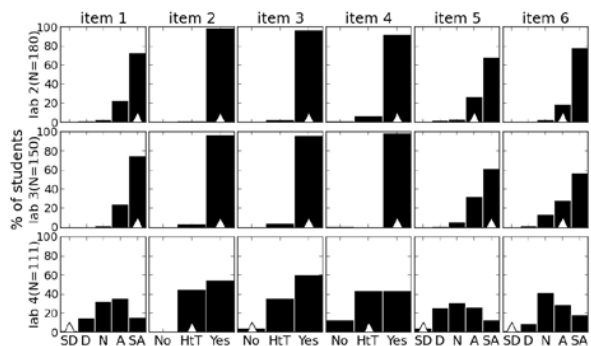

**Figure 1.** Distribution of student ratings in summer 2013. In each subplot, the horizontal axis indicates the grading options available (where "SD" to "SA" stands for "Strongly Disagree" to "Strongly Agree", and "HtT" stands for "Hard to Tell"). The white triangle indicates the instructors' rating on the same item in the same video.

## Rating after Intervention

Figure 2 compares the distribution of student ratings between Summer 2013 and Fall 2013. Figure 2 suggests that after intervention was provided, the distribution of students' ratings significantly shifted toward the lower end on most items in Lab 4. In particular, students in fall appeared to have adopted more rigorous evaluative norms for what constitutes an introduction (item 2). While in Summer 2013 55% of the students believed that the video in Lab 4 had introduced the problem and stated the main results, the percentage of students who believed so dropped by half in the fall. Examining students' written comments in Fall 2013, we found that many students pointed out that "*[the video] did not really explain the project*" and/or that "*[the video] does not include brief results in the introduction*" as the major issue of the Lab 4 video on this evaluated item. Similar significant shifts in the distribution of student rating on the same item was found in Lab 3 as well. Examination of student comments suggests that 63% of the students who rated the Lab 3 video as "hard to tell" or "no" on item 2 in Fall 2013 pointed out that the statement of the main result was missing in the introduction of this video. While the instructors did not necessarily take points off for this item on the hidden calibration video, the importance of providing a summary of the main result was constantly emphasized in the instructor comments for all the practice videos. It appears that with the intervention provided in Fall 2013, more students considered this as an important aspect in their evaluation of the quality of an introduction.

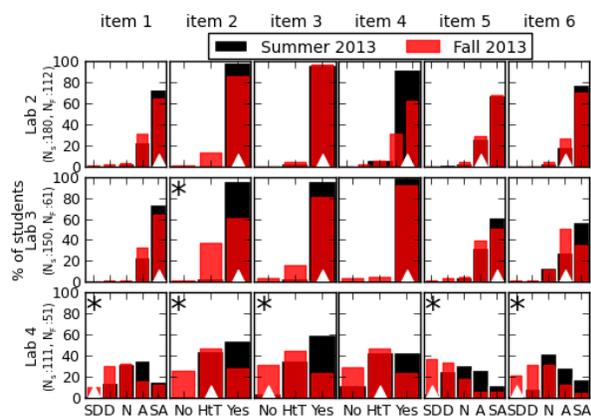

**Figure 2.** Distribution of student ratings between Summer 2013 and Fall 2013. $N_S$ and $N_F$ indicate the number of students in summer and fall, respectively. Cases where the distribution changes significantly (with p-value <0.05 in a two-sample Kolmogorov-Smirnov test) are indicated by an asterisk at the upper left corner. In fall 2013, a 5-point scale was used in lab 2 for item 4. If student responses on this item were binned into a 3 point scale by combining two extreme options on each side, there was no significant difference between student evaluation in summer and fall.

Examination of student ratings on Lab 4, item 2 suggests that students had also shifted their views on what constitutes a discussion of the physics principles involved. While in Summer 2013 60% of the students believed that the physics principles involved were identified in the Lab 4 video, in the fall, only 31% of the students thought so. Moreover, the percentage of students who believed that the physics principles were not discussed at all increased from 5% to 24%. One student in fall explicitly pointed out that "*You showed nothing concerning the Energy principle nor Newton's 2nd Law. Of course you showed how you implemented the formulas, but some explanation on the relevant physics is from my point of view a crucial part of a lab report.*" Similar statements were more commonly found in Fall 2013 than in Summer 2013, suggesting that students in fall took up an evaluative norm on this item that is more rigorous and more similar to that of the instructors.

In addition to the changes in items 2 and 3, the last row in Fig. 2 suggests that students might have also raised their bars on production quality (item 1), discussion of results (item 5) and the overall physics discussion (item 6). Moreover, for items 5 and 6, we also notice that students seemed more likely to pick up on instructor evaluative norms regarding severe and obviously apparent mistakes (e.g., when the video was rated strongly negative by instructors). When the flaws in the video were more nuanced, students did not seem to pick up on them as much. For example, on items 5 and 6, both Lab 2 and Lab 4 videos received an instructor rating lower than "Strongly Agree". However, the major issues associated with the Lab 4 video were severe; the presenter simply showed what he did without any explanation or discussion of the physics behind it. On the other hand, the flaws in the Lab 2 video were more nuanced: although the video presents a fair understanding of physics overall, the discussion was sometimes vague, involving small mistakes, and could have had more depth. Figure 2 shows that students' ratings in Fall 2013 lowered significantly in Lab 4, but not in Lab 2. It is possible that students in Fall 2013 may not necessarily consider those nuanced issues as crucial in their rating, and/or it might require more expertise in physics to be able to identify the issues in lab 2.

## CONCLUDING REMARKS

In this paper, we discussed an investigation into students' peer evaluation of video lab reports in an introductory physics MOOC by focusing on students' evaluative norms and how these norms were affected by the different interventions provided to students. Our analysis suggests that with only a single, early intervention, students adopt a lenient grading standard. After the practice videos were introduced giving students continued practice, students picked up more rigorous evaluative norms from instructors and were more critical in their assessment of others' lab reports. Our analysis also suggests that most of the instructor evaluative norms that students picked up in Fall 2013 were those that were less physics-intensive and more focused on issues of presentation (e.g., the lack of discussion). When the videos under evaluation involved issues in their physics discussion that were more nuanced (e.g., not enough depth or vague statements), students were less likely to pick them out. In future work, we plan to explore the mechanism for why and how students take up evaluative norms differently to help shape the design of a more effective peer evaluation system.

This work was supported by the Bill and Melinda Gates Foundation.

## REFERENCES

1. L. Pappano, The Year of the MOOC – NY Times (2012, November 2)
2. J. Aiken, et al., 2013 *PERC Proc.*, 53-56 (2014)
3. E. Price, et al., 2012 *PERC Proc.*, 318-321 (2013)
4. D. Brown, Tracker, ver. 7.51 (2013)
5. D. Scherer, et al., VPython, ver. 6.05 (2013)
6. R. K. *Thornton, e*t **al**., *Am. J. Phys.* **66**, 338 (1998)